# Multiscales and cascade in isotropic turbulence

Zheng Ran

Shanghai Institute of Applied Mathematics and Mechanics,Shanghai University, Shanghai 200072, P.R.China

The central problem of fully developed turbulence is the energy cascading process. It has revisited all attempts at a full physical understanding or mathematical formulation. The main reason for this failure are related to the large hierarchy of scales involved, the highly nonlinear character inherent in the Navier-Stokes equations, and the spatial intermittency of the dynamically active regions. Richardson has described the interplay between large and small scales and the phenomena so described are known as the Richardson cascade. This local interplay also forms the basis of a theory by Kolmogorov. In this letter, we use the explicit map method to analyze the nonlinear dynamical behavior for cascade in isotropic turbulence. This deductive scale analysis is shown to provide the first visual evidence of the celebrated Richardson cascade, and reveals in particular its multiscale character. The results also indicate that the energy cascading process has remarkable similarities with the deterministic construction rules of the logistic map. Cascade of period-doubling bifurcations have been seen in this isotropic turbulent systems that exhibit chaotic behavior. The ' cascade ' appears as an infinite sequence of period-doubling bifurcations.

The general pattern of turbulent motion can be described (according to Taylor and Richardson ) in the following way. The mean flow is accompanied by turbulent fluctuations imposed on it and having different scales, beginning with maximal scales of the order of the 'external scale- $L$ ' of turbulence to the smallest scales of the order of the distance $h$ at which the effect of viscosity becomes appreciable (the internal scale of turbulence). Most large scale fluctuations receive energy from the mean flow and transfer it to fluctuations of smaller scales. Thus there appears to be a flux of energy transferred continuously from fluctuations of large scales to those of smaller scales. Dissipation of energy, that is, transformation of energy into heat, occurs mainly in fluctuations of scale $h$. The amount of energy $e$ dissipated in unit time per unit volume is the basic characteristic of turbulent motion for all scales. Richardson's hypothesis says that, in a turbulent flow, energy is continually passed down from the large-scale structures to small scales, where it is destroyed by viscous stresses. Moreover, this is a multi-stage process involving a hierarchy of vortex sizes. Kolmogorov's theory, on the other hand, asserts that the statistical properties of the small scales depend only on and on the rate at which energy is passed down the energy cascade. Update, it appears that these statements cannot be formally 'proven' in any deductive way[1-4] .

In this letter, we will re-examine Richardson's idea of an energy cascade, in which, it is claimed, energy is passed down from large to small scales by a repeated sequence of discrete steps. But way should energy always pass from large to small scales, and way must this involve a distributed hierarchy of eddy sizes ? Is it not possible for energy to travel upscale ? Or perhaps we can transfer energy directly from large eddies to the Kolmogorov micro scale in one single action, thus by passing the cascade. In this letter, we shall show that Richardson's energy cascade is a direct consequence of vortex period-doubling bifurcation, based on the new scaling equation we found before. Now the Richardson picture is not entirely implausible, but it does raise at least two fundamental questions. First, is there some generic process which causes energy always to pass from large to small scale and not from small to large ? Second, why must the process be a multistage one ?

For complete self-preserving isotropic turbulence, the corresponding scaling equation takes the form of Eq.(1) [5-10]. We can see that this is a closed non-linear equation for the length scale $l(t)$.

$$\frac{d^2 l}{dt^2} + \frac{(2a_1 + a_2)n}{2l^2}\frac{dl}{dt} - \frac{a_1 a_2 n^2}{2l^3} = 0 \tag{1}$$

We are only interested in the special solutions in the following form:

$$l(t) = l_0 (t + t_0)^{\frac{1}{2}} \tag{2}$$

where $l_0$ are parameters to be chosen. We also know that the solutions of the turbulence scaling equation can be classified into two different kinds: P-model and N-model, and their solutions read as follows

$$l_P = \sqrt{2a_1 n(t + t_0)} \tag{3}$$

$$l_N = \sqrt{a_2 n(t + t_0)} \tag{4}$$

From the small dynamics theory based on our recent work [10], the parameter and its variable range are given as

$$s = \frac{a_2}{2a_1} \tag{5}$$

$$s_m = \frac{2}{3} + m, \quad m = 0,1,2,...M + 1 \tag{6}$$

A useful dimensional system must be comprised of a number of fundamental (base) entries (dimensions) that are sufficient to define the magnitude of any numerically expressible quantity. These fundamental dimensions may be chosen rather arbitrarily, but for practical reasons, should be chosen appropriately. The selection of a dimensional system must be carried out in two steps. The first step is to select the number of fundamental dimensions, and second is to select the standard magnititudes for these dimensions.

For a turbulent scale system, if we choose the fundamental scale as:

$$l_{ref} = l_P \equiv \sqrt{a_2 n(t + t_0)} \tag{7}$$

This analysis leads a natural description of the cascade which is a multistage process, involving a hierarchy of vortices of varying size. Here, it is conventional to talk of these different size structures as eddies, which conjures up a picture of spherical-like objects of different diameters, denoted as $\{l_m, m = 0,1,...,M+1\}$, and

$$l_m = \sqrt{\frac{3}{2+3m}} \cdot l_{ref} \tag{8}$$

The $l_m$ represents a hierarchy of eddy sizes from the integral scale $L$ down to $h$. It should be noted that for all $m$, $l_m$ is a decreasing function on $m$. Then the order in volume from one generation to next is:

$$L = l_0 > l_1 > l_2 ... > l_{m-1} > l_m > l_{m+1} > .... l_M > l_{M+1} = h \tag{9}$$

Let $F_m$ be defined as follows:

$$F_m = \left(\frac{l_{ref}}{l_m}\right)^2 \tag{10}$$

It follows from above analysis that for each $m = 0,1,2,...,M,M+1$ straightforward manipulation on $F_m$ using the definition of Eq.(10) show that $F_m$ are necessarily linked by a simple recursion relation:

$$F_{m-1} - 2F_m + F_{m+1} = 0 \tag{11}$$

This relation is valid for any 'scaling system'. However, for present discussion on turbulent cascade, the whole solution of $F_m$ must satisfy two boundaries conditions, where $F_0$ and $F_{M+1}$ are fixed. These two boundary conditions uniquely determine the solution, and the final expression of $F_m$ is then

$$F_m = \frac{M+1-m}{M+1} \cdot F_0 + \frac{m}{M+1} \cdot F_{M+1} \tag{12}$$

It is natural to ask what their interpretation is in the present framework for $l_m$ and $l_{m+1}$. We would like to argue that Eq.(12) may be rewritten in the symbolic form

$$l_{m+1}^2 = \frac{l_m^2}{1 + A \cdot l_m^2} \tag{13}$$

where

$$A \equiv \frac{1}{M+1}\left[\frac{1}{h^2} - \frac{1}{L^2}\right]. \tag{14}$$

Accordingly, if the fundamental scale is chosen as follows

$$l_R = \sqrt{2a_1 n(t+t_0)} \tag{15}$$

At last we have another expression for the scale

$$q_{m+1}^2 = q_m^2 + B \tag{16}$$

where

$$B \equiv \frac{1}{M+1} \cdot L^2 - \frac{1}{M+1} \cdot h^2 \tag{17}$$

At this stage, we have obtained two different expressions of the turbulent scales. Physically, the true values of the turbulent scale area will be the arithmetic mean of these two different scales areas, here, we have

$$y_m \equiv \frac{l_m^2 + q_m^2}{2} \tag{18}$$

We can convert this into a difference equation only for $y_m$, in view of the assumption of $A \ll 1$:

$$y_{m+1} = -2Ay_m^2 + y_m + \frac{1}{2}B \tag{19}$$

The map (19) is an instance of logistic map. Here, we could as well call it the logistic map of isotropic turbulence. In the further investigation, in fact, we would not use the original form of eq. (19). Moreover, we would like use the standard form which is used in the chaos theory by the definition of topologically conjugate

$$x_{m+1} = ax_m(1-x_m) \tag{20}$$

It is easy to show that

$$a = 1 + \sqrt{1-4c} \tag{21}$$

$$c = -2 + \frac{9}{2} \cdot \left(\frac{R}{R_c}\right)^{-\frac{3}{2}} - \frac{9}{4} \cdot \left(\frac{R}{R_c}\right)^{-3} \tag{22}$$

Note that the control parameter $a$ only depends on the relative Reynolds number. $x_m$ could be regarded as the mean area of the Taylor micro scale in isotropic turbulence.

Computer experiment turns out to be loaded with marvelous scientific discoveries. We want to explore the behavior of the quadratic iterate (20) for all values of the parameter $a$ between $1$ to $4$. Figure 1. shows that the results for all parameter $a$. We note that for parameter $a > 3$ the final state is not a mere point but a collection of 2,4 or more points. For $a = 4$, of course we have the chaos, and the points of the final state densely fill up the complete interval. Sometimes this image is also called the Feigenbaum diagram. Indeed, this diagram is a remarkable fractal, and later we will see that it is closely related to the famous Mandelbrot set.

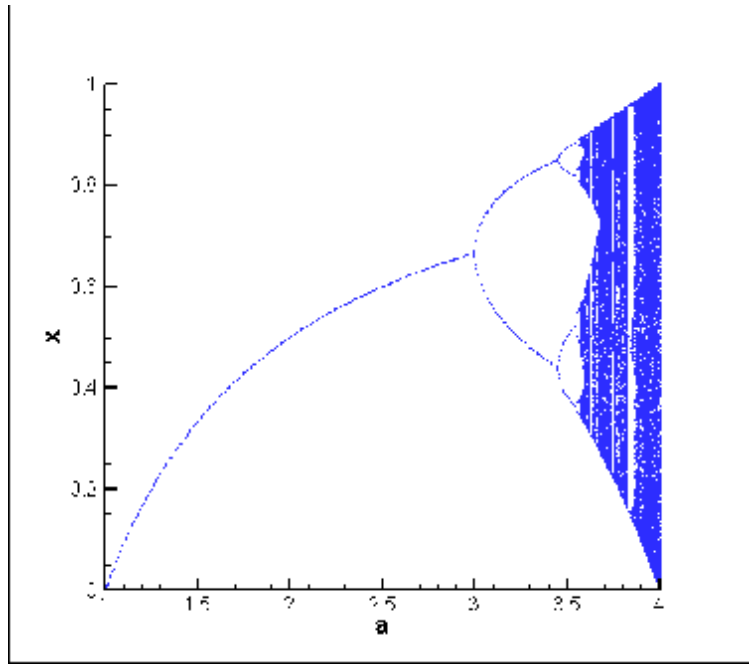

**Figure 1. The Feigenbaum diagram of isotropic turbulence**

One essential structure seen in the Feigenbaum diagram above is that of a branching tree which portrays the qualitative changes in the dynamical behavior of the iterate (20). Out of a major stem we see two branches bifurcating and out of these branches we see two branches bifurcating again, and then two branches bifurcating out of these again, and so on. This is period-doubling regime of the scenario. The Feigenbaum diagram has self-similarity features that are both of a qualitative nature and a quantitative one. The qualitative features are best analyzed through the methodology of fractal geometry. The structure of self-similarity properties, which, means that the route from order to chaos is one with infinite detail and complexity. This covers main features of the classical turbulence cascade picture due to Richardson.

Eq.(20) is also the well-known logistic equation. The equation was originally proposed for the description of the dynamics of a population of organisms that appear in discrete generations, such as insects. May present a number of examples ranging from genetic problems to sociology which are modeled by equate ion of the type (20). The logistic equation is one of the most often discussed prototypes of the complex behavior of deterministic systems. The best visible is a sequence of successive period doublings. It seems that periodic points first appear in order $1, 2, 4, 8, \ldots, 2^n, \ldots$ occur as $a$ increases. Specifically, let $\dfrac{R_m}{R_c}$ denote the value of bifurcation where a $2^n - $ cycle first appears. Then table.1 reveals that the critical bifurcation values of further period-doubling. Note that the successive bifurcations come faster and faster. Ultimately the bifurcation parameter converges to a limiting value $a_\infty$. For $a > a_\infty$, the orbit diagram reveals an unexpected mixture of order and chaos, with periodic windows interspersed between chaotic clouds of dots. This implies the transition from laminar to turbulence.

**TABLE.1 The critical Reynolds number for period doubling bifurcation**

| $m$ | | $\dfrac{R_m}{R_c}$ |
|---|---|---|
| 1 | $1 \to 2$ | 2.080083823 |
| 2 | $2 \to 2^2$ | 3.096733934 |
| 3 | $2^2 \to 2^3$ | 3.511518701 |
| 4 | $2^3 \to 2^4$ | 3.619876484 |
| 5 | $2^4 \to 2^5$ | 3.644190644 |
| 6 | $2^5 \to 2^6$ | 3.649451759 |
| … | … | … |
| $\infty$ | $2^\infty \to$ chaos | 3.650890633 |

One of the main goals in the development of theory of chaotic dynamical system has been to make progress in understanding of turbulence. The attempts to related turbulence to chaotic motion got strong impetus from the celebrated paper by Ruelle and Takens . Considerable success has been achieved mainly in the area: the onset of turbulence. For fully developed turbulence, many questions remain unanswered. The aim of this letter is to show that there are dynamical systems that are much simpler than the Navier-Stokes equations but that can still have turbulent states and for which many concepts developed in the theory of dynamical systems can be successfully applied. In this connection we advocate a broader use of the universal properties of a wide range of isotropic turbulence phenomena. Even for the case of fully developed turbulence, which contains an extreme range of relevant length scales, it is possible, by using the present model, to reproduce a surprising variety of relevant features, such as multifractal cascade, intermittency. This letter reverts to possible applications of the Navier-Stokes equations to studies of the nature of turbulence.

ACKNOWLEDGMENTS. The work was supported by the National Natural Science Foundation of China (Grant Nos.90816013, 10572083).